\documentclass[conference,10pt,twocolumn]{IEEEtran}
\normalsize
\usepackage{amsmath, amsthm, amssymb}
\usepackage{siunitx}
\usepackage{filecontents}
\usepackage[numbers,sort&compress]{natbib}
\usepackage{epsfig}
\usepackage{latexsym}
\usepackage{epstopdf}
\usepackage{multirow}
\usepackage{graphicx}
\usepackage[caption=false]{subfig}
\usepackage{algorithm}
\usepackage{algpseudocode}
\usepackage{subfig}
\usepackage{float}

\begin{document}

\title{Coded Network Function Virtualization: Fault Tolerance via In-Network Coding
}
\author{
    \IEEEauthorblockN{A. Al-Shuwaili\IEEEauthorrefmark{1}, O. Simeone\IEEEauthorrefmark{1}, J. Kliewer\IEEEauthorrefmark{1} and P. Popovski\IEEEauthorrefmark{2}}
    \IEEEauthorblockA{\IEEEauthorrefmark{1}CWiP, Department of Electrical and Computer Engineering, New Jersey Institute of Technology, Newark, NJ, USA.
    \\\{ana24,  osvaldo.simeone, and jkliewer\}@njit.edu}
    \IEEEauthorblockA{\IEEEauthorrefmark{2}Department of Electronic Systems,
Aalborg University, Aalborg, Denmark.
    \\\{petarp\}@es.aau.dk}
}
\maketitle

\begin{abstract}
Network Function Virtualization (NFV) prescribes the instantiation of network functions on general-purpose network devices, such as servers and switches. While yielding a more flexible and cost-effective network architecture, NFV is potentially limited by the fact that commercial off-the-shelf hardware is less reliable than the dedicated network elements used in conventional cellular deployments. The typical solution for this problem is to duplicate network functions across geographically distributed hardware in order to ensure diversity. In contrast, this letter proposes to leverage channel coding in order to enhance the robustness on NFV to hardware failure. The proposed approach targets the network function of uplink channel decoding, and builds on the algebraic structure of the encoded data frames in order to perform in-network coding on the signals to be processed at different servers. The key principles underlying the proposed coded NFV approach are presented for a simple embodiment and extensions are discussed. Numerical results demonstrate the potential gains obtained with the proposed scheme as compared to the conventional diversity-based fault-tolerant scheme in terms of error probability.

\end{abstract}


\section{introduction}
 Network Function Virtualization (NFV) is a novel architectural paradigm for cellular wireless networks that has been put forth within the European Telecommunications Standards Institute (ETSI) with the goal of simplifying network management, update and operation \cite{ets}. NFV decouples the Network Functions (NFs), such as baseband processing at the base stations and firewalling or routing at the core network, from the physical network equipment on which they run. This is done by leveraging virtualization technology in order to map NFs into Virtual Network Functions (VNFs) that are instantiated on Commercial Off-The-Shelf (COTS) hardware resources, such as servers, storage devices and switches \cite{ref1,rel}. NFV enables an adaptive ``slicing'' of the available network physical resources so as to accommodate different network services, e.g., mobile broadband, machine-type or ultra-reliable communications \cite{ref2}. 
\begin{figure}[!t] \label{fig1}
        \centering
        \includegraphics[width=\columnwidth]{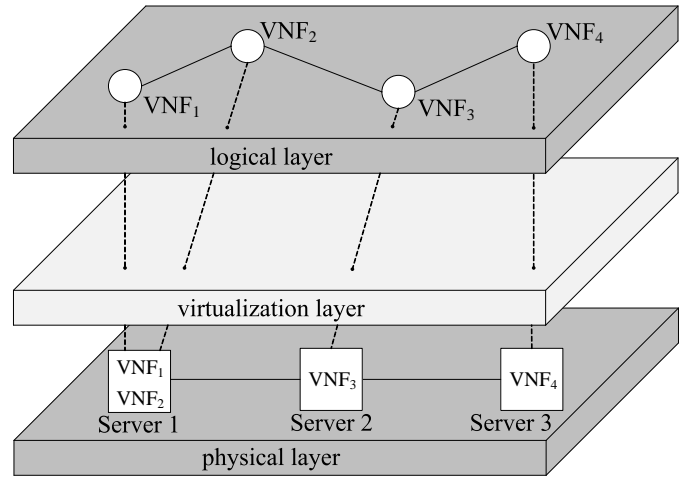}
        \caption{{\small{Simplified architectural view of NFV.}}}
        \label{sys}
\end{figure}

A simplified view of the NFV architecture is illustrated in Fig. 1 \cite{etsi, ref1, ref2, rel}. The top layer in this architecture describes the logical functionality of the given network service as a so called forwarding graph, which characterizes the functional relationship among the VNFs that implement the network service. The bottom layer contains the general-purpose hardware appliances that provide storage, computation and networking capabilities. Finally, the intermediate virtualization layer  is responsible for mapping VNFs to physical resources. In the example of Fig. 1, VNF1 and VNF2 are instantiated on  Virtual Machines (VMs) running on Server 1, while VNF3 and VNF4 are instantiated on VMs running on Server 2 and Server 3, respectively. The combination of the last two layers constitutes the Network Function Virtualization Infrastructure (NFVI), and the three planes are under the control of a Network Functions Virtualization Management and Orchestration (NFV-MANO) block (see \cite{ref1, rel, ets} for details). Most research activity on NFV focuses on the design of mapping rules between VNFs and hardware resources via the solution of mixed integer problems (see, e.g., \cite{rami}). 
 
One of the key challenges for the adoption and deployment of NFV is the fact that COTS hardware is significantly less reliable than the dedicated network devices used in conventional network deployments \cite[Sec. VI]{rel}. Hardware outages may in fact be caused by random failures, intentional attacks, software malfunction or disasters. This problem is motivating an emerging line of work, also within ETSI, on developing fault-tolerant virtualization strategies for NFV \cite{etsi,etsi2,4}. The typical solution, as summarized in \cite{etsi2}, is to adapt to NFV well established policies introduced in the context of virtualization for data centers. These strategies are based on \textit{overprovisioning} and \textit{diversity}: NFs are split into multiple constituents VNFs, which are then mapped on VMs instantiated on multiple distributed servers in order to minimize the probability of a disruptive failure as well as the mean time to recovery from a failure.

In this work, we propose a novel principle for the design of fault-tolerant NFV that moves from \emph{diversity-based} solutions to \emph{coded} solutions.  The proposed approach addresses the NF of uplink data decoding in a Cloud Radio Access Network (C-RAN) architecture, in which the baseband processing operations of the base station are carried out remotely at the ``cloud'' \cite{aldo}. The focus on uplink channel decoding is dictated by the fact that the latter is known to be among the most demanding baseband functions in terms of computational complexity (see, e.g., \cite{rost,comsoc}). 

The proposed coded NFV solution leverages the algebraic structure of the transmitted coded data frames in order to enhance the robustness of channel decoding. To elaborate, assume that there are a number of servers on which VMs carrying out channel decoding can be instantiated, as illustrated in Fig. 2. A conventional diversity-based technique would duplicate the decoding task at multiple servers. In this letter, instead, we propose a coded approach, whereby received data frames are encoded prior to being processed by the VMs that implement decoding at the distributed servers. This letter elaborates on a simple embodiment of this idea, which is illustrated in Fig. 3 and introduced in Sec. III after a description of the system model in Sec. II. Numerical examples are provided in Sec. IV. Extensions and more general applications of the principle of coded NFV are presented in Sec. V, while the concluding remarks are given in Sec. VI. We finally observe that the idea presented here is related to the concepts of coded computations put forth in \cite{ali,5}, but, to the best of the authors' knowledge, the idea of performing coding to robustify the operation of NFV is first proposed here.
\begin{figure}[!t] \label{fig2}
        \centering
        \includegraphics[width=\columnwidth]{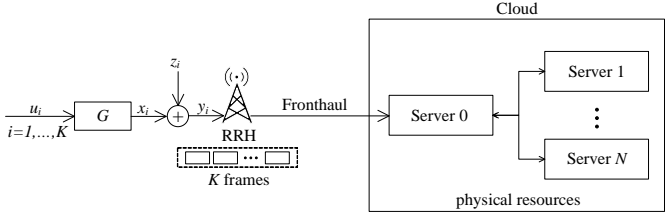}
        \caption{{\small{ NFV set-up for the virtualization of the NF of uplink channel decoding. Server 0 acts as controller and is assumed to be reliable, while servers $1,\ldots,N$ can carry out the computationally heavy NF of channel decoding but may be unavailable, due to failures or overload, with probability $q$.}}}
        \label{sys}
\end{figure}

\section {System Model } 
 We consider a C-RAN system implemented by means of NFV, and focus on the implementation of the NF of uplink channel decoding. In this system, as illustrated in Fig. 2, a Remote Radio Head (RRH) is connected to the cloud by means of a fronthaul link. The RRH forwards the received baseband packets to the cloud on the fronthaul link in order to enable channel decoding. We assume the overprovisioning of hardware resources, such that $N$ servers are available in the cloud, each of which can run a VM performing the decoding of a single received frame in the allotted time. More specifically, due to latency constraints, the decoding of $K$ received data frames should be carried out on the servers by allocating at most one frame to decode to each of $N\geq K$ servers. As in conventional implementations (see, e.g., \cite{etsi2}), we further assume that the VMs implementing channel decoding on Servers $1, \ldots,N$ are managed by a controller VM, which is characterized by lower computational requirements and is instantiated at a server, marked as Server $0$ in Fig. 2, that is connected with bidirectional links to Servers $1,\ldots, N$. 
 
 Servers $1,\ldots, N$ are distributed strategically across multiple locations throughout the service provider's network, and are hence assumed to have independent availabilities \cite{etsi2}. In particular, we assume that each one of Servers $1,\ldots,N$ fails independently with probability $q$. It is emphasized that a failure here means that a server is not available to perform the given task within an acceptable deadline due to software or hardware issues (see Sec. I).

The transmitted frames are encoded with the same $(n,k)$ linear code with a given rate $k/n$, such as convolutional, turbo or LDPC codes.  Furthermore, in order to present the key ideas, we consider first a Binary Symmetric Channel (BSC) model between the user under study and the RRH. As a result, for each transmitted frame $x_i \in \{0,1\}^n$, with $i \in \{1,\ldots,K\}$, the transmitted signal can be written as $x_i=u_i G$, where $u_i \in \{0,1\}^k$ represents the data encoded in the $i$th frame and $G$ is the $k \times n$ generator matrix of the code. Furthermore, the signal received for the $i$th frame is given as $y_i=x_i \oplus z_i$ where $z_i$ is a vector of independent Bernoulli variables with probability $p$ of being equal to 1. Generalizations of the system model are discussed in Sec. V. 

Throughout, we take as the performance metric of interest the probability of error, that is, the complement of the probability that decoding of all the $K$ frames is carried out successfully by the cloud. Note that, according to the introduced model, a failure may occur due to either errors on the communication channel between user and RRH or due to a failure of the servers.

\section {Fault tolerance via coded NFV } \label{sec:2}
In this section, we first review the conventional diversity-based fault-tolerant approach as applied to the problem at hand of uplink channel decoding in a C-RAN via NFV. We then present the proposed coded NFV approach. For both schemes, we focus on the case $N=3$ and $K=2$ and present a simple analysis of the probability of error. The problem statement in the general case is treated in Section V.

 \subsection {Fault Tolerance via Diversity}
 
 A conventional solution based on diversity is illustrated in Fig. 3(a) for $N=3$ servers and $K=2$ frames. In this scheme, the controller VM instantiated at Server 0 duplicates one of the received frames, namely $y_2$ in the figure, at the input of both Server 2 and Server 3. Server 1, Server 2 and Server 3 each run  a VM that performs channel decoding as well as error detection (via a Cyclic Redundancy Check test) on the input frame. The outcome of the decoders is fed back to the VM in Server 0. We note that, for general values of $N$ and $K$ with $N>K$, the scheme would just duplicate one or more frames at the input of multiple servers.


\begin{figure}[htp]
\centering
\subfloat[Diversity-based scheme]{%
  \includegraphics[clip,width=\columnwidth]{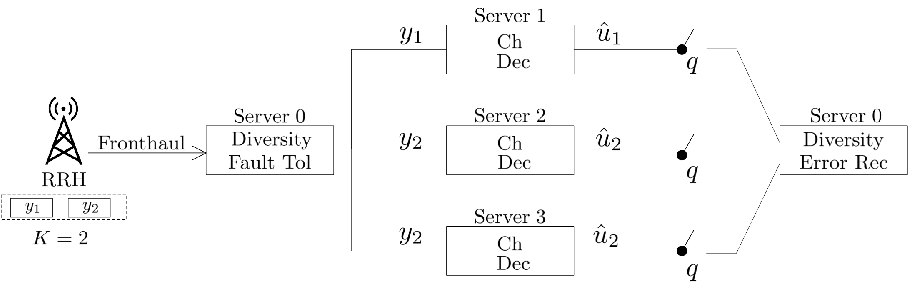}%
}

\subfloat[Coded NFV scheme]{%
  \includegraphics[clip,width=\columnwidth]{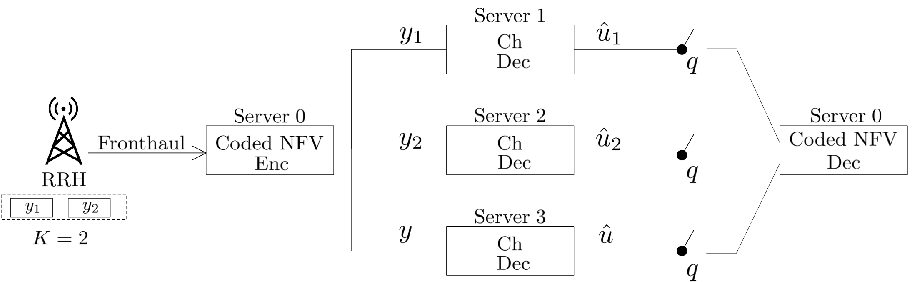}%
}

\caption{\small{ Illustration of the idea of coded NFV for channel decoding in the case of an overprovisioning factor of $N/K=3/2$ ($N=3$ servers  for $K=2$ received frames): (a) Mapping of NF and servers for a conventional diversity-based scheme in which one of the received frames is duplicated at the input of two servers; (b) Mapping of NF and servers for the proposed coded NFV scheme in which the XOR of the two received frames $y = y_1 \oplus y_2 $ is the input to Server 3, whose output is $\hat{u} = \hat{u}_1 \oplus \hat{u}_2 $.}}

\end{figure}
 The conventional diversity-based system succeeds in decoding both packets as long as: (\emph{i}) Server 1 decodes correctly data $u_1$ and is available; and (\emph{ii}) Server 2 and/or Server 3 decode correctly $u_2$ and are available. As a consequence, the error probability can be written as
    \begin{equation}
 \text{P}_{\text {err}} = 1- \sum_{\substack{{\mathcal{S} \subseteq \{1,2,3\}: }\\ |\mathcal{S}|\geq 2, \{1\}\in \mathcal{S}}} \text{Pr}(\mathcal{S}) (1-q)^{|\mathcal{S}|},
 \end{equation}
where $|\mathcal{S}|$ is the cardinality of set $\mathcal{S}$, which is a subset of the Servers 1, 2 and 3; $\text{Pr}(\mathcal{S})$ is the probability that only the decoders in $\mathcal{S}$  successfully decode the input frame, while the rest of the servers decode incorrectly.

\subsection {Fault Tolerance via Coded NFV}
In the proposed coded NFV scheme, as illustrated in Fig. 3(b), Server 0 pre-processes the received frames by computing the linear combination $y=y_1 \oplus y_2$ of the received frames $y_1$ and $y_2$. Note that this operation is of much lower complexity as compared to channel decoding. Server 0 then assigns frame $y_1$ for decoding at Server 1, frame $y_2$ to Server 2 and frame $y$ to Server 3. A key observation is that Server 3 can decode over the same linear code as Server 1 and 2 since we have 
\begin{equation}
y = y_1 \oplus y_2 = (u_1 \oplus u_2) G \oplus (z_1\oplus z_2).
\end{equation}
Hence, Server 3 can decode $u=u_1\oplus u_2$ over a BSC with parameter $2p(1-p)$, which is the probability that the effective noise $z_1\oplus z_2$ equals 1. As a result, as long as \emph{any} two servers decode successfully and are available, Server 0, which receives the outputs of all other servers as in the diversity-based scheme, can decode both data messages $u_1$ and $u_2$.

Based on the description above, the proposed coded NFV scheme  can be interpreted as a form of \emph{concatenated code} in which the outer linear code encodes each frame, while the inner NFV code is applied on the noisy received signals in order to obtain robustness with respect to infrastructure failures. The probability of error  for this scheme is given by 
\begin{equation}
 \text{P}_{\text {err}} =  \sum_{\substack{{\mathcal{S} \subseteq \{1,2,3\}: }\\ |\mathcal{S}|\geq 2}} \text{Pr}(\mathcal{S}) (1-q)^{|\mathcal{S}|},
 \end{equation}
where $\text{Pr}(\mathcal{S})$ is defined as in (1), with the key difference that Server 3 decodes based on (2).

We conclude this section by emphasizing that, beside the advantages  in terms of the error probability which will be further discussed in Sec. IV, the proposed coded scheme increases the Minimum Failure Removal (MFR) \cite{4}. The MFR is the minimum number of servers whose removal leads to  failure. In particular, with the conventional diversity-based scheme, even the non availability of a single server, namely Server 1, causes a failure, while the proposed scheme has a MFR of two.

\section{Numerical Results}
In this section, we present numerical experiments to compare the performance of the conventional diversity-based scheme and the proposed coded NFV for the  presented example with $N=3$ and $K=2$. To this end, we consider a $1/2$ feedforward convolutional code, in which the constraint length is 7, the code generator polynomial matrix is $[171 \;\; 133]$, with $k=70$ and $n=140$ and Viterbi decoders are implemented at Servers 1, 2 and 3. We evaluate the probabilities $\text{Pr}(\mathcal{S})$ in (1) and (3) via Monte Carlo simulations.

 The error probability as a function of the servers' failure probability $q$ is plotted in Fig. 4 for the indicated values of the BSC parameter $p$ for both schemes. It is seen that, in the regime in which hardware failures have similar or smaller probability as compared to channel errors, coded NFV can provide significant gains. For instance, to achieve $\text{P}_{\text{err}} \approx 2\times10^{-3}$ with $p=0.05$, the conventional diversity-based method  requires hardware with a server failure probability of $q = 10^{-4}$, while the coded NFV  requires $q \approx 10^{-3}$, which is an order of magnitude larger. 
    

\begin{figure}[!t] \label{fig2}
        \centering
        \includegraphics[width=\columnwidth]{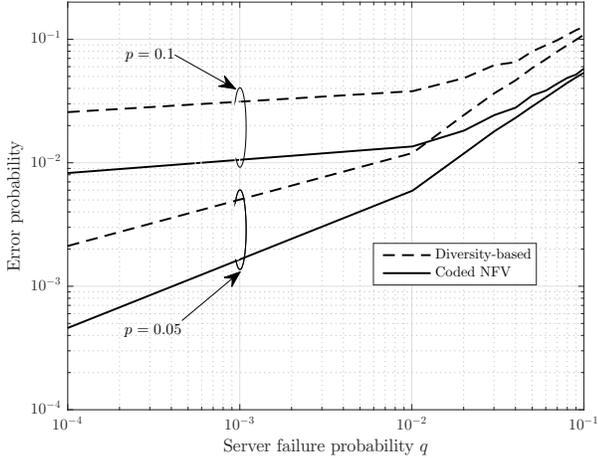}
        \caption{{\small{ Error probability versus the server failure probability $q$ for diversity-based and coded NFV schemes $(n=140, k=70,  N=3, K=2)$.}}}
        \label{sys}
\end{figure}

\section {Extensions}
The robust coded NFV scheme was presented in Sec. III-B for $N=3$, $K=2$  and for a BSC channel between user and RRH. In this section, we briefly discuss extensions.

1) \textit{Frames encoded with different rates}: Different rates can be accommodated by using rate-compatible codes obtained from the same master linear code for each frame.

2) \textit{Additive Gaussian noise or fading channels}: For such channels between user and RRH, lattice codes can be used to encode the frames, instead of  linear binary codes. The input to Server $3$ (see Fig. 3(b)) is computed as the sum of the received signals on the real field, or complex field for complex Gaussian or fading channels. Server 3 then decodes the XOR of the two messages, namely $u=u_1 \oplus u_2$, by decoding over the lattice code using the technique of computation over a multiple access channel (see \cite{6} and references therein for an introduction).

3) \textit{Generalization to any values of the parameters $N$ and $K$}: For any values of $N$ and $K$, each one of the $n$ bits input to Server $j$, with $j=1,\ldots,N$, is obtained as a binary-field linear combination of the corresponding bits of the received frames $y_i$, with $i=1,\ldots,K$. The resulting NFV code can be then described by a $K\times N$ generator matrix $G_{\text{NFV}}$ such that the input bits to the servers can be computed as $y G_{\text{NFV}} $, with $y \in \{0,1\}^K$ collecting one bit from every frame. Note that $G_\text{NFV}=\bigl( \begin{smallmatrix} 
  1 & 0 & 1\\
  0 & 1 & 1 
\end{smallmatrix} \bigr)
$ in the discussed $N=3$, $K=2$ example.


Regarding the design of the generator matrix $G_{\text{NFV}}$, we note that an NFV code benefits from a sparse structure of $G_{\text{NFV}}$, as well as from a large minimum distance ---  two conflicting requirements. For the former requirement, as seen in Sec. III-B, summing more received signals
at the input of a server increases the noise level. 
More formally, the NFV code operates over an erasure channel in which the probability of error
associated to each server $k$ equals $q+(1-q)f(d_k)$, where $d_k$ is
the number of ones in the $k$th column of matrix $G_{\text{NFV}}$ and $f(d)$
represents the probability of incorrect decoding when $d$ received
signals are summed (which is an increasing function of $d$). This novel property of NFV codes sets an interesting
research challenge for code design.

\section{concluding remarks}
Software-based virtual network functions enabled by NFV are less reliable than those provided via the traditional hardware-based platforms. To alleviate this shortcoming, this letter proposed to enhance traditional diversity-based solutions by means of channel coding. The proposed solutions addresses the important network function of uplink channel decoding at the base station and leverages the algebraic structure of the received encoded data frames. Open questions for future work encompass the design of NFV codes and the application of the principle of coded NFV to other network functions, such as routing.

\bibliographystyle{ieeetran}
\bibliography{rr}
\end {document}